\documentclass[12pt]{article}

\usepackage{graphicx} 

\begin{document}

\title{Modified frequentist determination of confidence intervals for Poisson distribution}    

\author{S.I.Bitioukov and N.V.Krasnikov
\\
INR RAS, Moscow  117312}

\maketitle

\begin{abstract}
We propose  modified frequentist definitions for the determination of confidence intervals 
for the case of Poisson statistics. We require that
$\displaystyle 1-\beta^{'} \geq \sum_{n=o}^{n_{obs}+k} P(n|\lambda) \geq \alpha^{'} $. 
We show that this definition is equivalent to
the Bayesian method with prior $\pi(\lambda) \sim \lambda^{k}$.
Other generalizations are also considered. In particular, we propose modified symmetric 
frequentist definition which corresponds to the Bayes approach with the prior 
function  $\displaystyle \pi(\lambda) \sim \frac{1}{2}(1 + \frac{n_{obs}}{\lambda})$.
Modified frequentist definitions for the case of nonzero background 
are proposed. 

%\keywords{Confidence interval; Poisson distribution.}
\end{abstract}

%\ccode{PACS Nos.: 02.50.-r.} 

\newpage
\section{Introduction}
In high energy physics one of the standard problems \cite{1} is the determination of the 
confidence intervals for the parameter 
$\lambda$ in  Poisson distribution
\begin{equation}
P(n|\lambda) = \frac{\lambda^n}{n!}\exp(-\lambda)\,.
\end{equation} 
There are two methods to solve this problem - the frequentist and the Bayesian.

In Bayesian method  \cite{1,2} due to Bayes theorem 

$$P(A|B) = \frac{P(B|A)P(A)}{P(B)} $$

the probability density for the $\lambda$ parameter 
is determined as
\begin{equation}
p(\lambda |n_{obs}) = \frac{P(n_{obs} |\lambda)\pi(\lambda)}{\int_{0}^{\infty} (P(n_{obs} |\lambda^{'})
\pi(\lambda^{'})d \lambda^{'}} \,.
\end{equation}
Here $\pi(\lambda)$ is the prior function and in general it is not known that is the 
main problem of the Bayesian method. Formula (2) reduces the statistics problem to the 
probability problem. At the  ($1 -\alpha$) probability level the parameters 
$\lambda_{up}$ and $\lambda_{down}$ are determined from the equation \footnote{Usually 
$\alpha$ is taken equal to $0.05$.}
\begin{equation}
\int_{\lambda_{down}}^{\lambda_{up}} p(\lambda|n_{obs}) d \lambda = 1 -\alpha
\end{equation} 
and the unknown  parameter $\lambda$ lies between $\lambda_{down}$ and $\lambda _{up}$ 
with the probability $1 - \alpha$. 
The solution of the equation (3) is not unique. 
One can define 
\begin{equation}
\int_{\lambda_{up}}^{\infty} p(\lambda|n_{obs}) d \lambda = \alpha^{'} \,,
\end{equation}
\begin{equation}
\int_{0}^{\lambda_{down}} p(\lambda|n_{obs}) d \lambda = \beta^{'} \,.
\end{equation}
In general the  parameters $\alpha^{'}$ and $\beta^{'}$ are arbitrary except 
the evident equality
\begin{equation}
\alpha^{'} + \beta^{'} = \alpha\,.
\end{equation}

The most popular are the following options \cite{1}:

1. $\lambda_{down} = 0$ - upper limit.

2. $\lambda_{up} = \infty$ - lower limit.

3. $\int_{0}^{\lambda_{down}} p(\lambda|n_{obs}) d \lambda = 
\int_{\lambda_{up}}^{\infty} p(\lambda|n_{obs}) d \lambda = \frac{\alpha}{2}$ - symmetric interval.

4. The shortest interval -  $p(\lambda|n_{obs})$ inside the interval is bigger or equal to  
$p(\lambda|n_{obs})$ outside the interval.

In frequentist approach the Neyman belt construction~\cite{3}~(see Fig.~1~\cite{PDG}) 
is used for the determination of the confidence intervals.

\begin{figure}[!Hhtb]
%  \begin{center}
\includegraphics[width=0.90\textwidth]{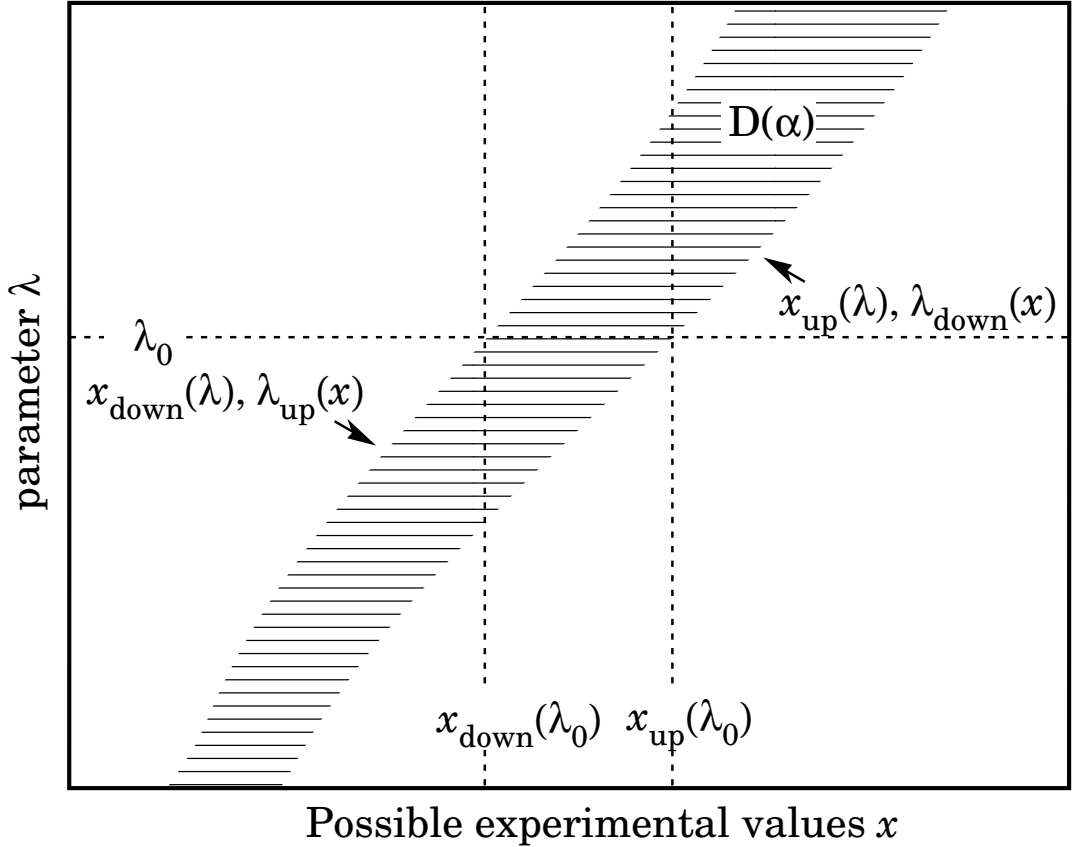}   
\caption{Neyman belt construction.}
    \label{fig:1} 
%  \end{center}
\end{figure}

%\begin{figure}[!Hhtb]
%  \begin{center}
%\includegraphics[width=0.90\textwidth]{fig01.eps}   
%\caption{Neyman belt construction}
%    \label{fig:2} 
%%  \end{center}
%\end{figure}

For the continuous observable $-\infty < x  < \infty $ with the probability density 
$f(x, \lambda)$ \footnote{Here $\lambda$ is some unknown parameter and $\int^{\infty}_{-\infty}
f(x,\lambda)dx = 1$.} we require that
\begin{equation}
\int^{x_{up}}_{x_{down}}f(x,\lambda)dx = 1 - \alpha\,,
\end{equation}
or
\begin{equation}
\int^{\infty}_{x_{up}}f(x,\lambda)dx = \beta^`\,,
\end{equation}
\begin{equation}
\int^{x_{down}}_{-\infty}f(x,\lambda)dx = \alpha^`\,,
\end{equation}
\begin{equation}
\alpha^` + \beta^` = \alpha\,.
\end{equation}
The equations\footnote{Here $x_{obs}$ is the observed value of random variable $x$.}
\begin{equation}
\int^{\infty}_{x_{obs}}f(x,\lambda_{down})dx = \beta^`\,,
\end{equation}
\begin{equation}
\int^{x_{obs}}_{-\infty}f(x,\lambda_{up})dx = \alpha^`\,
\end{equation}
determine the interval 
of possible values $\lambda_{down} \leq \lambda \leq \lambda_{up}$ of the 
parameter $\lambda$ at the $(1 - \alpha)$ confidence level.

For Poisson distribution $P(n|\lambda)$ the analog of the equation (7) has the form
\begin{equation}
\sum_{n_{down}(\lambda)}^{n_{up}(\lambda)} P(n|\lambda) \geq 1 - \alpha \,.
\end{equation}

The equations for the determination of $\lambda_{down}$ and $\lambda_{up}$ (analogs 
of the equations (11, 12 ) ) have the form  \cite{4,5,6}
\begin{equation}
\sum_{n=n_{obs}}^{\infty}P(n|\lambda_{down}) = \beta^{'}\,,
\end{equation}
\begin{equation}
\sum_{n=0}^{n_{obs}} P(n|\lambda_{up}) = \alpha^{'}\,.
\end{equation}

As a consequence of the equations (14, 15) we find that for  $\lambda_{up} = \lambda_{down}$ 
the probability
$1 - \alpha^{'} - \beta^{'} = -  P(n_{obs}|\lambda_{up}) < 0 $ that contradicts to our intuition that 
the probability 
$P(\lambda_{down} \leq \lambda \leq \lambda_{up})= 1 - \alpha^` - \beta^` \rightarrow 0$
for    $\lambda_{down}\rightarrow \lambda_{up}$ , i.e. $\alpha^{'} + \beta^{'} \rightarrow 1 $. 
For the case of continuous random variable $x$ with smooth probability 
density $f(x, \lambda)$ as a consequence of the equations (11,12) for 
 $\lambda_{down}\rightarrow \lambda_{up}$ the evident limit $\alpha^` + \beta^` \rightarrow 1$ 
takes place.
 
In this paper~\footnote{The main results of this paper are contained in ref.~\cite{BitKras}.}  
we propose  the modified frequentist definitions of  confidence interval 
for the case of Poisson distribution. We show that the modified frequentist definitions are  
equivalent to the Bayesian approach. The organization of the paper is the following. 
In Section 2 we propose modified frequentist definitions  of confidence inteval 
and show its equvalence to 
the Bayes method. In Section 3 we discuss the case of nonzero background. 
Section 4 contains concluding remarks.

\section{Modified frequentist definitions of the confidence interval}

For the case of continuous random variable $x$
the equations (11,12) are equivalent to the equations
\begin{equation}
\int^{\infty}_{x_{obs}}f(x,\lambda_{down})dx = \beta^` \,,
\end{equation}
\begin{equation}
\int^{\infty}_{x_{obs}}f(x,\lambda_{up})dx = 1 -\alpha^` \,
\end{equation}
or to the  equations 
\begin{equation}
\int^{x_{obs}}_{-\infty}f(x,\lambda_{up})dx = \alpha^`\,,
\end{equation}
\begin{equation}
\int^{x_{obs}}_{-\infty}f(x,\lambda_{down})dx = 1 - \beta^`\,.
\end{equation}

One can find that the inequalities

\begin{equation}
1 - \beta^{'} \geq \int^{x_{obs}}_{-\infty}f(x,\lambda)dx \geq \alpha^`\,
\end{equation}

and

\begin{equation}
1 - \alpha^` \geq \int^{\infty}_{x_{obs}}f(x,\lambda)dx \geq  \beta^` \,
\end{equation}
are equivalent and they determine the interval of possible values 
 $\lambda_{down} \leq \lambda \leq \lambda_{up}$ 
(see eqs.(11,12 )) at the $(1 - \alpha)$ confidence level.

For Poisson distribution $P(n|\lambda)$ in closed  analogy 
with the inequalities (20,21) we require that \footnote{We can consider the inequalities (22,23) as 
modified frequentist definitions for the determination of confidence intervals.} 
 \begin{equation}
1-\beta^{'} \geq P_{-}(n_{obs}|\lambda) \geq \alpha^{'} \,
\end{equation}
or
 \begin{equation}
1-\alpha^{'} \geq P_{+}(n_{obs}|\lambda) \geq \beta^{'} \,,
\end{equation}
where
\begin{equation}
P_{-}(n_{obs}|\lambda) = \sum_{n=0}^{n_{obs}}P(n|\lambda) \,,
\end{equation}
\begin{equation}
P_{+}(n_{obs}|\lambda) = \sum_{n=n_{obs}}^{\infty}P(n|\lambda) \,.
\end{equation}

For Poisson distribution the inequalities (22) and (23) lead to the equations 
\begin{equation}
P_{-}(n_{obs}|\lambda_{down}) = 1 - \beta^{'} \,,
\end{equation}
\begin{equation}
P_{-}(n_{obs}|\lambda_{up}) = \alpha^{'} \,
\end{equation}
and 

\begin{equation}
P_{+}(n_{obs}|\lambda_{down}) = \beta^{'} \,,
\end{equation}
\begin{equation}
P_{+}(n_{obs}|\lambda_{up}) = 1 - \alpha^{'} \,
\end{equation}
for the determination of $\lambda_{down}$ and $\lambda_{up}$. 
As we mentioned  before the choice of
$\lambda_{down}$ and  $\lambda_{up}$ is not unique. Probably the most natural choice is the use of 
the ordering principle. According to this principle 
 we require  that the probability density $P(n_{obs}|\lambda)$ inside 
the confidence interval $[\lambda_{down}, \lambda_{up}]$ is bigger or equal to 
the probability density outside this interval. For Poisson distribution this requirement 
leads to the formula
\begin{equation}
P(n_{obs}|\lambda_{down}) = P(n_{obs}|\lambda_{up})\,
\end{equation}
for the determination of $\lambda_{up}$ and $\lambda_{down}$.
For such ordering principle $\alpha^{'}$ and $\beta^{'}$ are not independent quantities. It is 
natural to use $\alpha = \alpha^{'} + \beta^{'}$ as a single free parameter.

Unlike to the case of continuous variable 
the equations (14, 15), (26, 27) and (28, 29) are not equivalent for the discrete variable $n$ 
and they differ in the presence or absence of $P(n_{obs}| \lambda_{up,down})$ in some equations.
For instance, for $\beta^` = 0$, $\alpha^` = \alpha$ (upper limit case) the equations (15) and 
(27) 
coincide and read as 
\begin{equation}
\sum^{n_{obs}}_{n=0} P(n| \lambda_{up}) = \alpha \,,
\end{equation}
while the equation (29) is equivalent to
\begin{equation}
\sum^{n_{obs}-1}_{n=0} P(n| \lambda_{up}) = \alpha \,,
\end{equation}
For $n_{obs} = 3$ and $\alpha = 0.05$  we find that 
\begin{equation}
\lambda \leq 7.75\,,(Eq.~31)\,,
\end{equation} 
\begin{equation}
\lambda \leq 6.30\,,(Eq.~32)\,.
\end{equation}

Due to the identity \cite{6}
\begin{equation}
P_{-}(n_{obs}|\lambda) = \int_{\lambda}^{\infty}P(n_{obs}|\lambda^{'})d\lambda^{'}
\end{equation}
the confidence interval $[\lambda_{down}, \lambda_{up}]$ for the modified frequentist definition 
(22)  is determined from the equations
\begin{equation}
\alpha^{'} = \int_{\lambda_{up}}^{\infty}  P(n_{obs}|\lambda^{'}) d\lambda^{'}\,,
\end{equation}
\begin{equation}
\beta^{'} = \int_{0}^{\lambda_{down}}P(n_{obs}|\lambda^{'})d\lambda^{'}\,.
\end{equation}
The   parameter $\lambda$ lies in the interval
\begin{equation}
\lambda_{down} \leq \lambda \leq \lambda_{up}
\end{equation} 
with the probability $(1- \alpha^{'} - \beta^{'})$.
So we see that our modified frequentist definition (22) is equivalent to Bayes 
definitions (3, 4, 5) with flat prior $\pi(\lambda) = 1$, namely:
\begin{equation}
\int_{\lambda_{down}}^{\lambda_{up}} P(n_{obs}|\lambda^{'})d\lambda^{'} = 1 - \alpha^{'} - \beta^{'} \,.
\end{equation}
One can show that our modified frequentist definition (23) (eqs. (28,29)) 
is equivalent to the Bayes 
approach with  the prior function  $\pi(\lambda) \sim \frac{1}{\lambda}$.

The coverage of the definition (22) means the following. For a hypothetical 
ensemble of similar experiments the probability to observe the number of events 
$n\leq  n_{obs}$ satisfies the inequalities (22).

Note that the  equations  for the determination of an upper limit 
$\lambda_{up}$ in frequentist and modified frequentist approach  (22) coincide whereas 
the equations for the determination of lower limit are different. 
Namely, the equation (26) is equivalent 
to the equation 
\begin{equation}
\sum_{n=n_{obs}+1}^{\infty}P(n|\lambda_{down}) = \beta^{'}\,.
\end{equation}

Classical frequentist equation (15) for the determination of $\lambda_{up}$ is equivalent 
to Bayes equation (4) with flat prior while the equation (14) for the determination of 
$\lambda_{down}$  
is equivalent to the Bayes equation (5) with the prior 
$\pi(\lambda) \sim \frac{1}{\lambda}$.

It is possible to generalize our modified frequentist definition (22), namely:
\begin{equation}
1 - \beta^{'}   \geq P_{-}(n_{obs}|\lambda;k)  \geq \alpha^{'}\,,
\end{equation}
where
\begin{equation}
P_{-}(n_{obs}|\lambda;k) \equiv \sum _{n = 0}^{n_{obs} +k}P(n|\lambda)
\end{equation}
and  $k = 0, {\pm}1, {\pm} 2, ...$

One can find that definition  (41) leads to Bayes equations (4, 5) with the 
prior function $\pi(\lambda) \sim \lambda^{k}$. 
The cases $k = 0$ and $k = -1$ are equivalent to the inequalities (22) and (23). 
Upper limits for three values of $k = 0, \pm 1$ 
are shown in Table~\ref{tab:1} ($\alpha = 0.1$), 
in Table~\ref{tab:2}  ($\alpha = 0.05$) and, correspondingly, 
in Fig.~2 and Fig.~3.

\begin{table}[h]
\begin{center}
\caption
%\tbl
{Upper limits ($\lambda_{up}$) for confidence level 90\% ($\alpha = 0.1$).}
%{\begin{tabular}{@{}rrrr@{}} \toprule 
\begin{tabular}{|r|r|r|r|}
\hline
$n_{obs}$ &  k=-1 & k=0 & k=+1  \\ 
\hline
    0  &   -   &  2.30 &  3.89 \\
    1  &  2.30 &  3.89 &  5.32 \\
    2  &  3.89 &  5.32 &  6.68 \\
    3  &  5.32 &  6.68 &  7.99 \\
    4  &  6.68 &  7.99 &  9.27 \\
    5  &  7.99 &  9.27 & 10.53 \\
    6  &  9.27 & 10.53 & 11.77 \\
    7  & 10.53 & 11.77 & 12.99 \\
    8  & 11.77 & 12.99 & 14.21 \\
    9  & 12.99 & 14.21 & 15.41 \\
   10  & 14.21 & 15.41 & 16.60 \\
\hline
\end{tabular}
\end{center}
\label{tab:1}
%}
\end{table}
    
\begin{table}[h]
\begin{center}
\caption
%\tbl
{Upper limits ($\lambda_{up}$) for confidence level 95\% ($\alpha=0.05$).}
%{\begin{tabular}{@{}rrrr@{}} \toprule 
\begin{tabular}{|r|r|r|r|}
\hline
$n_{obs}$ &  k=-1 & k=0 & k=+1  \\ 
\hline
    0  &   -   &  3.00 &  4.74 \\
    1  &  3.00 &  4.74 &  6.30 \\
    2  &  4.74 &  6.30 &  7.75 \\
    3  &  6.30 &  7.75 &  9.15 \\
    4  &  7.75 &  9.15 & 10.51 \\
    5  &  9.15 & 10.51 & 11.84 \\
    6  & 10.51 & 11.84 & 13.15 \\
    7  & 11.84 & 13.15 & 14.43 \\
    8  & 13.15 & 14.43 & 15.71 \\
    9  & 14.43 & 15.71 & 16.96 \\
   10  & 15.71 & 16.96 & 18.21 \\  
%\botrule
\hline
\end{tabular}
\end{center}
\label{tab:2}
%}
\end{table}

\newpage
\begin{figure}[!Hhtb]
%  \begin{center}
\includegraphics[width=0.90\textwidth]{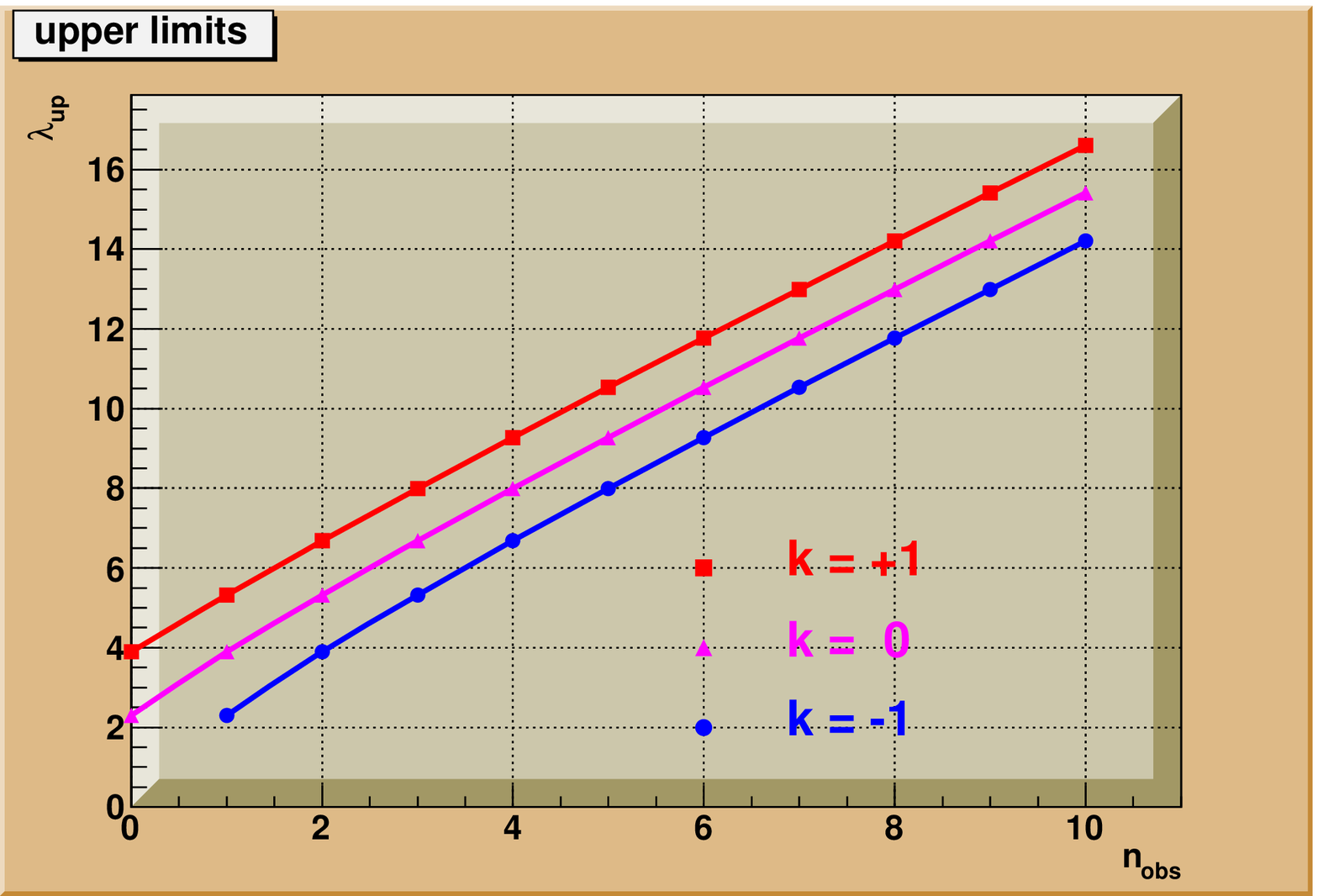}   
\caption{Upper limits ($\lambda_{up}$) for confidence level 90\% ($\alpha = 0.1$), $k=-1,~0,+1$.}
    \label{fig:2} 
%  \end{center}
\end{figure}

\begin{figure}[!Hhtb]
%  \begin{center}
\includegraphics[width=0.90\textwidth]{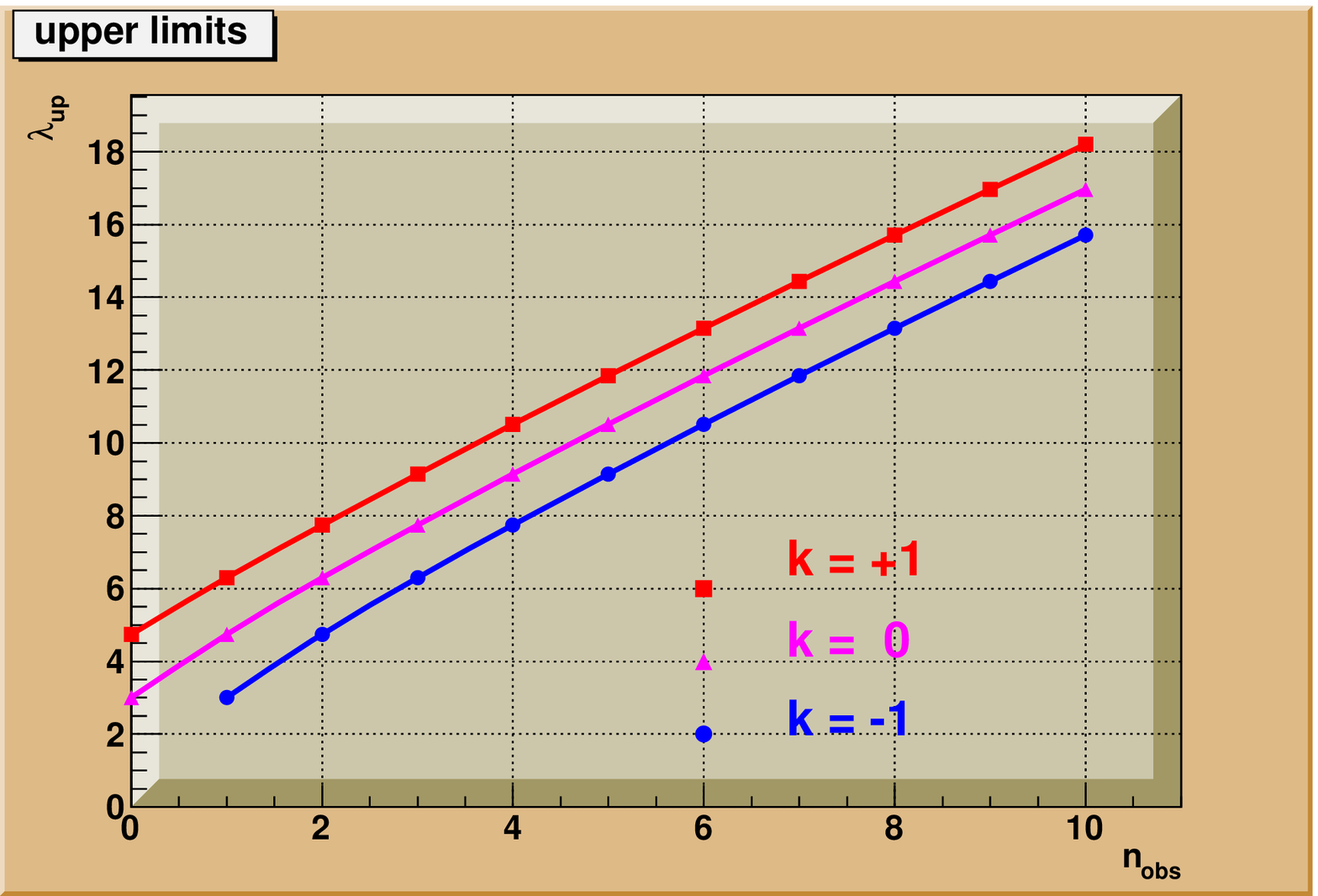}   
\caption{Upper limits ($\lambda_{up}$) for confidence level 95\% ($\alpha = 0.05$), $k=-1,~0,+1$.}
    \label{fig:3} 
%  \end{center}
\end{figure}

We can further generalize definitions (41, 42) by the introduction 
\begin{equation}
P_{-}(n_{obs}|\lambda; c_{k}) \equiv \sum_{k}c^2_{k}P_{-}(n_{obs}|\lambda;k) \,,
\end{equation}
where $\sum_{k} c^2_{k} = 1$.
 Again we require that
\begin{equation}
1 - \beta^{'} \geq P_{-}(n_{obs}|\lambda ;c_{k}) \geq \alpha^{'}\,.
\end{equation}
One can find that our definition (43, 44) is equivalent to Bayes approach with prior function
\begin{equation}
\pi(\lambda) \sim \sum_{k} c^2_{k}l_{k} \lambda^{k} \,,
\end{equation}
where
\begin{equation}
l_{k} = \frac{n!}{(n+k)!}\,.
\end{equation}

Note that in modified frequentist inequalities (22, 23) the term $P(n_{obs}|\lambda$ 
contributes in (22) and (23) that leads to nonequivalence of these inequalities.
One of the possible  symmetric generalizations of the modified frequentist inequalities 
(22,23) looks as follows
\begin{equation}
1-\beta^{'} \geq P_{-}(n_{obs}|\lambda) - \frac{1}{2}P(n_{obs}|\lambda) \geq \alpha^{'} \,,
\end{equation}
 \begin{equation}
1-\alpha^{'} \geq P_{+}(n_{obs}|\lambda)  - \frac{1}{2}P(n_{obs}|\lambda)       \geq \beta^{'} \,.
\end{equation}
The inequalities (47) and (48) are equivalent to each other and moreover they are 
equivalent to the Bayes approach with the prior function

\begin{equation}
\pi(\lambda) \sim \frac{1}{2}(1 + \frac{n_{obs}}{\lambda})\,.
\label{eq:our-prior}
\end{equation}

Upper limits for the prior~(\ref{eq:our-prior}) and for the Jeffreys prior~\cite{Jeffreys} 
$\pi(\lambda)\sim\frac{1}{\sqrt{\lambda}}$ can be found in Table~\ref{tab:3} ($\alpha = 0.1$) 
and in Table~\ref{tab:4} ($\alpha = 0.05$). 

\begin{table}
\begin{center}
\caption
%\tbl
{Upper limits ($\lambda_{up}$) for confidence level 90\% ($\alpha=0.1$).} 
%{\begin{tabular}{@{}rrrr@{}} \toprule  
\begin{tabular}{|r|r|r|}
\hline
$n_{obs}$ &  $\pi(\lambda)\sim\frac{1}{2}(1 + \frac{n_{obs}}{\lambda})$ & $\pi(\lambda)\sim\frac{1}{\sqrt{\lambda}}$ \\ 
\hline
    0  &  2.30 &  1.35  \\
    1  &  3.27 &  3.12  \\
    2  &  4.72 &  4.61  \\
    3  &  6.10 &  6.00  \\
    4  &  7.57 &  7.34  \\
    5  &  8.71 &  8.63  \\
    6  &  9.97 &  9.90  \\
    7  & 11.21 & 11.15  \\
    8  & 12.44 & 12.38  \\
    9  & 13.65 & 13.60  \\
   10  & 14.85 & 14.80  \\
\hline
\end{tabular}
\end{center}
\label{tab:3}
%}
\end{table}

\begin{table}
\begin{center}
\caption
%\tbl
{Upper limits ($\lambda_{up}$) for confidence level 95\% ($\alpha=0.05$).} 
%{\begin{tabular}{@{}rrrr@{}} \toprule 
\begin{tabular}{|r|r|r|}
\hline
$n_{obs}$ &  $\pi(\lambda)\sim\frac{1}{2}(1 + \frac{n_{obs}}{\lambda})$ & $\pi(\lambda)\sim\frac{1}{\sqrt{\lambda}}$ \\ 
\hline
    0  &  3.00 &  1.92  \\
    1  &  4.11 &  3.90  \\
    2  &  5.68 &  5.53  \\
    3  &  7.16 &  7.03  \\
    4  &  8.57 &  8.45  \\
    5  &  9.93 &  9.83  \\
    6  & 11.27 & 11.18  \\
    7  & 12.58 & 12.49  \\
    8  & 13.87 & 13.79  \\
    9  & 15.14 & 15.07  \\
   10  & 16.40 & 16.33  \\
\hline
\end{tabular}
\end{center}
\label{tab:4}
%}
\end{table}

%\begin{equation}
%P_{-1}(n_{obs}|\lambda_{up}) = \alpha^{'} \,,
%\end{equation}
%\begin{equation}
%P_{+1}(n_{obs}|\lambda_{down}) = \beta^{'} \,,
%\end{equation}
%where
%\begin{equation}
%P_{-1}(n_{obs}|\lambda) = \sum_{n=0}^{n_{obs}-1}P(n|\lambda) ~+ \frac{1}{2}P(n_{obs}|\lambda)\,,
%\end{equation}
%\begin{equation}
%P_{+1}(n_{obs}|\lambda) = \sum_{n_{obs}+1}^{n=\infty}P(n|\lambda) ~+ \frac{1}{2}P(n_{obs}|\lambda)\,%.
%\end{equation}
%Note that
%\begin{equation}
%P_{-1}(n_{obs}|\lambda) +  P_{+1}(n_{obs}|\lambda) = 1
%\end{equation}
%and $\alpha^{'} +\beta^{'} = 1$ for $\lambda_{up} = \lambda_{down}$.

\section{The case of nonzero background}

For  nonzero background the 
parameter $\lambda$ is represented 
in the form
\begin{equation}
\lambda = b + s \,.
\end{equation}
Here $b \geq 0$   is known background  and $s$ is unknown signal. 
In Bayes approach the generalization of the formula (2) reads
\begin{equation}
p(s |n_{obs},b) = \frac{P(n_{obs} |b+s)\pi(b,s)}{\int_{0}^{\infty}{P(n_{obs} |b+s^{'}) 
\pi(b,s^{'})d s^{'}}} \,.
\end{equation}
For flat prior we find
\begin{equation}
p(s |n_{obs},b) = \frac{P(n_{obs} |b+s)}{\int_{b}^{\infty} {P(n_{obs} |\lambda^{'})
d \lambda^{'}}} \,.
\end{equation}
The main effect of nonzero background is the appearance 
of the factor 
\begin{equation}
K(n_{obs},b) = \int_{b}^{\infty} P(n_{obs} |\lambda^{'})
d \lambda^{'}  
\end{equation}
in the denominator of the formula (52). For zero background 
$K(n_{obs}, b =0) = 1$.  One can interpret the appearance of additional factor 
$K(n_{obs},b)$ in terms of conditional probability. Really, for flat prior the 
$P(n_{obs},\lambda)d\lambda $ is the probability that parameter $\lambda$ lies 
in the interval $[\lambda, \lambda +d\lambda]$. For the case of nonzero  background 
$b$ parameter $\lambda = b + s \geq b$.
The probability that $\lambda \geq b$ is equal to $p(\lambda \geq b|n_{obs}) =
K(n_{obs},b)$. The conditional probability that $\lambda$ lies in the interval 
$[\lambda, \lambda +d\lambda]$ provided $\lambda \geq b$ is determined by 
the standard formula 
\begin{equation}
p(\lambda,n_{obs}|\lambda \geq b)d\lambda = 
\frac{P(n_{obs}|\lambda)d\lambda}{p(\lambda \geq b)}  = 
\frac{P(n_{obs}|\lambda)d\lambda}{K(n_{obs} s)}
\end{equation} 
and it coincides with the Bayes formula (52).

In the frequentist approach the naive  generalization of the inequality (22) is
\begin{equation}
1-\beta^{'} \geq P_{-}(n_{obs}|s +b) \geq \alpha^{'} \,.
\end{equation}
One can show that 
\begin{equation}
1 - \alpha^{'} - \beta^{'} = \int^{b+s_{up}}_{b+s_{down}}P(n_{obs}|\lambda^{'})d\lambda^{'} 
\leq \int^{\infty}_{b}P(n_{obs}|\lambda^{'})d\lambda^{'}\,.
\end{equation}
As a consequence of the inequality (56)
the probability that the signal $s$ lies 
in the interval $0 \leq s \leq \infty$ is  equal 
to $\int^{\infty}_{  b}P(n_{obs}|\lambda^{'})d\lambda^{'}$ and it 
is less than unity for nonzero background $b>0$ that contradicts to the intuition that 
the full   probability that the signal $s$ lies between
zero and infinity must be equal to unity. To cure this drawback let us require that
\footnote{The interpretation of the inequality (57) is as follows. We can consider 
the $P_{-}(n_{obs}|b)$ as the probability that $\lambda \geq b $.
The ratio  $ \frac{P_{-}(n_{obs}|s +b)}{P_{-}(n_{obs}|b)}$ is the conditional probability 
that $\lambda \geq b+s $ provided $\lambda \geq b $.}  
\begin{equation}
1-\beta^{'} \geq 
\frac{P_{-}(n_{obs}|s +b)}{P_{-}(n_{obs}|b)} 
\geq \alpha^{'} \,.
\end{equation}
The inequality (57) leads to the equations for the determination of $s_{down}$ and $s_{up}$ 
which coincide with the corresponding Bayes equations. The generalization of the inequalities 
(57) is straightforward, for instance the inequality (44) reads

\begin{equation}
1 - \beta^{'} \geq 
\frac{P_{-}(n_{obs}|b+s ;c_{k})}{P_{-}(n_{obs}|b ;c_{k})} 
\geq \alpha^{'} \,.
\end{equation}
Upper limit on the signal $s$ derived from the inequality (58) coincides with 
the upper limit in $CL_{s}$ method \cite{7,8}.

\section{Conclusions}

To conclude let us stress our main result. For Poisson distribution 
we have proposed modified frequentist definitions of the confidence interval and have shown 
the equivalence of the modified frequentist approach and Bayes approach. It means 
in particular that frequentist 
approach is  not unique.

 This work has been supported by RFBR grant N 10-02-00468Á.

\end{document}